\documentclass{aa}
\usepackage{graphicx}
\begin{document}

\authorrunning{K\"apyl\"a \& Korpi}
\titlerunning{ Resolution dependence of the Shakura--Sunyaev $\alpha$ }

   \title{Numerical study of the magnetorotational instability in
   weakly magnetised accretion disks}

   \subtitle{Resolution dependence of the Shakura--Sunyaev $\alpha$}

   \author{P. J. K\"apyl\"a
          \and
          M. J. Korpi
	  }

   \offprints{P. J. K\"apyl\"a\\
	  \email{Petri.Kapyla@cc.oulu.fi}
	  }

   \institute{Astronomy Division, Department of Physical Sciences,
              University of Oulu, PO BOX 3000, 90014 University of
              Oulu, Finland\\ }

   \date{Received ; accepted}

   \abstract{In this letter, we present numerical calculations made to
   investigate the possible resolution dependence of the Shakura \&
   Sunyaev (\cite{shasun}) viscosity parameter $\alpha$ from local
   magnetohydrodynamic simulations of the magnetorotational
   instability (MRI). We find that the values of $\alpha$ do indeed
   depend significantly on the numerical resolution but also that when
   the highest resolutions attainable by the computational resources
   available are used, the growth of the $\alpha$--parameter seems to
   saturate. The values of $\alpha$ are at most of the order of
   $10^{-3}$, which indicates that the sole presence of turbulence due
   to dynamo generated magnetic field in the disk is not enough to
   reproduce $\alpha$s of the order unity which could explain some
   observational results (e.g. Cannizzo \cite{canniz}).

   }

   \maketitle


\section{Introduction}

   The biggest problem in the theory of accretion disks has been the
   lack of knowledge of a widely applicable physical process which
   could reproduce the observed effective angular momentum
   transport. A decade ago, the magnetorotational instability was
   realised to have signifigance in the context of accretion disks
   (Balbus \& Hawley \cite{balhaw1}). The MRI was seen to excite and
   sustain turbulence giving rise to turbulent viscosity and outward
   angular momentum transport in local MHD calculations (e.g. Hawley
   et al. \cite{hgbI}, \cite{hgbII}; Brandenburg et al. \cite{bnstI},
   \cite{bnstII}; Stone et al. \cite{stone}).

   The efficiency of this process can be measured in terms of the
   Shakura \& Sunyaev $\alpha$--parameter which parametrises the
   turbulent viscosity as $\nu_{t} = \alpha c_{\rm s} H$, where
   $c_{\rm s}$ is the sound speed, and $H$ the height of the disk. The
   above parametrisation was made assuming the motion of the gas to be
   subsonic and that the turbulent eddies are smaller than the disk
   height, resulting in $\alpha$ as a dimensionless number whose
   values vary between zero and one. The interpretation of the
   lightcurves of some dwarf novae indicates that the value of
   $\alpha$ should be of the order $10^{-2}\!-\!1$ (e.g. Cannizzo
   \cite{canniz}).

   However, in the aforementioned numerical studies the values of
   $\alpha$ have been at least an order of magnitude too low to
   account for the observational results. On the other hand, the
   numerical resolution was seen to affect the values of $\alpha$
   (e.g. Brandenburg et al. \cite{bnstII}; Miller \& Stone
   \cite{millstone}). Doubling the resolution increased the value of
   $\alpha$ by a factor of $\sim\!1.5$ in these studies. This possible
   dependence has not been studied systematically yet, casting some
   doubt that high enough $\alpha$s can be obtained from local MRI
   calculations. Therefore, the purpose of this paper is to perform
   the previously missing survey on the resolution dependence of the
   $\alpha$--parameter.

\vspace*{-0.5cm}
\section{The MHD model}

   \begin{figure*}
   \centering
   \includegraphics[width=0.9\textwidth]{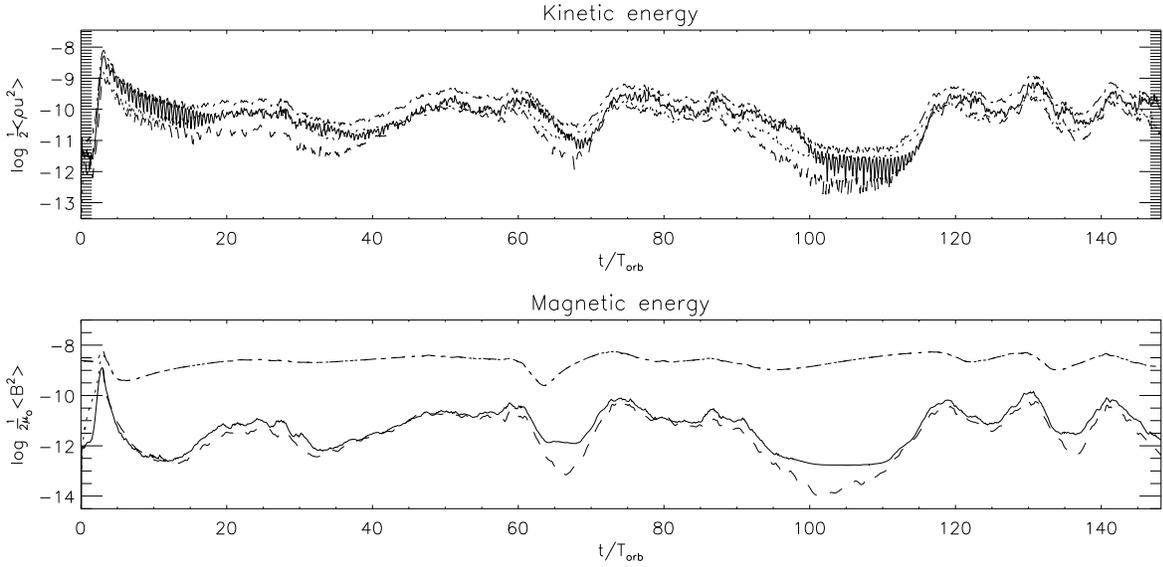}
   \vspace*{-4.2cm}
      \caption{Turbulent kinetic and magnetic energies from the run
      R1. The different components are denoted by solid (radial),
      dotted (azimuthal), and dashed (vertical) lines. The total
      energy is denoted by dash dotted line.}  
      \label{FigEne}
      \vspace*{-0.3cm}
    \end{figure*}

   The computational domain in our calculations is a rectangular box
   with dimensions $L_{i}$, $i = x, y, z$, at a radius $R_{0}$ from
   the centre of rotation. The distance to the centre of force is
   large in comparison to the box dimensions (i.e. $R_{0} \gg L_{i}$),
   under which assumption new locally cartesian coordinates can be
   taken into use and the equations of motion can be linearised
   (e.g. Spitzer \& Schwarzschild \cite{spischwa}). This
   simplification is usually referred to as the shearing box
   approximation, in which we solve a set of non--ideal MHD--equations
   in cartesian coordinates (see Caunt \& Korpi \cite{CauKo}).

   We apply small molecular viscosities everywhere and additionally
   artificial shock- and hyperviscosities in order to resolve shocks
   and (unphysical) small scale oscillations, respectively. The
   artificial viscosities also serve the purpose of stabilising the
   numerics.
 
   We adopt periodic boundary conditions in the radial and azimuthal
   directions, and closed insulating stress free boundaries in the
   vertical direction. The radial boundary takes into account the
   basic shear flow. Numerical implementation of the model described
   above is presented in the paper by Caunt \& Korpi (\cite{CauKo}).

   \begin{table}
      \caption[]{
      The values of maximum and time averaged $\alpha$--parameters 
      for the two sets of simulations. Time averages were taken up to
      $13\,T_{\rm orb}$ (duration of the largest calculation), where 
      $T_{\rm orb} = 2\pi/\Omega_{0}$ is the orbital period.}  
      \label{Calcu}
     $$  \vspace*{-0.45cm}
         \begin{array}{p{0.1\linewidth}ccc}
            \hline
            \noalign{\smallskip}
            Run      & \rm Resolution & \alpha_{\rm max} [10^{-3}] & <\!\alpha\!>_{\rm t} [10^{-4}]\\
            \noalign{\smallskip}
            \hline
            \noalign{\smallskip}
            T1 &  15 \times  15 \times  31 &  2 &  3 \\
            T2 &  31 \times  31 \times  63 & 13 & 10 \\
            T3 &  47 \times  47 \times  95 & 22 & 16 \\
            T4 &  63 \times  63 \times 127 & 23 & 32 \\
            T5 &  95 \times  95 \times 191 & 32 & 39 \\
            T6 & 127 \times 127 \times 255 & 28 & 41 \\
            \hline
            \noalign{\smallskip}
            R1 &  31 \times  31 \times  63 & 14 &  7 \\
            R2 &  47 \times  47 \times  95 & 16 & 18 \\
            R3 &  63 \times  63 \times 127 & 32 & 26 \\
            \noalign{\smallskip}
            \hline
         \end{array}
     $$ 
   \end{table}

   \begin{figure*}
   \centering
   \includegraphics[width=0.9\textwidth]{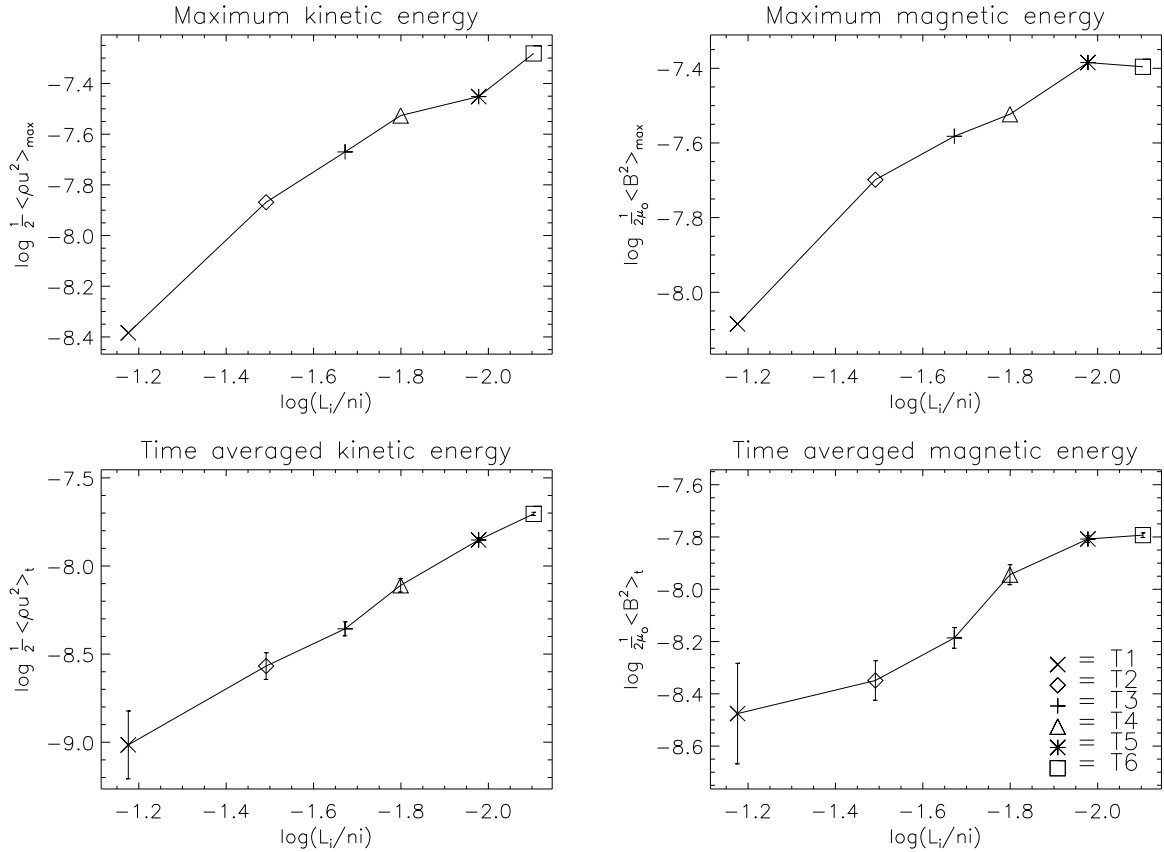}
      \vspace*{-0.35cm}
      \caption{
               Maximum and time averaged energies as a function of mesh
               size, $\Delta_{i} = L_{i}/ni$, where $ni$ is the number of 
               gridpoints and $i = x,y,$ or $z$. The errorbars are given
               by a constant factor divided by the number of
               datapoints used in the calculation.
               } 
      \label{ResEne}
      \vspace*{-0.3cm}
    \end{figure*}
\vspace*{-0.4cm}
\section{Physical setup}
   We adopt the same dimensionless quantities as Brandenburg et
   al. (\cite{bnstI}), i.e. length is measured in units of the initial
   density scale height $[x] = H_{0}$, time in units of $[t] =
   (GM/H_{0}^{3})^{-1/2}$, and the density in units of $[\rho] =
   \rho_{0}$, the initial density at the midplane of the disk. The
   unit of magnetic field will then be $[B] = [u](\mu_{0}
   \rho_{0})^{1/2}$, where $[u] = [x]/[t]$. Dimensionless quantities
   are obtained by prescribing
   \begin{eqnarray}
   H_{0} = GM = \rho_{0} = \mu_{0} = 1\,.
   \end{eqnarray}
   Initially the disk is in hydrostatic equilibrium which yields a
   Gaussian density stratification
   \begin{eqnarray}
   \ln \rho = \ln \rho_{0} - \frac{z^{2}}{H_{0}^{2}}\,.
   \end{eqnarray}
   The initial magnetic field configuration is a sinusoidally
   distributed vertical field in the radial direction giving a zero
   net field. In terms of the vector potential this configuration can
   be represented as
   \begin{eqnarray}
   \mathbf{A} = \frac{L_{x}}{2\pi} B_{0} \cos \Big(\frac{2\pi x}{L_{x}}\Big) \hat{\mathbf{y}}\,,
   \end{eqnarray}
   where $B_{0}$ is calculated from the plasma beta, $\beta = 
   2\mu_{0}p/B_{0}^{2}$, which has an initial value of 100 in 
   the midplane of the disk. Random velocity perturbations of the 
   order $10^{-3}c_{\rm s}$ were added in the velocity field to start
   up the instability.

\vspace*{-0.3cm}
\section{Calculations}
   Two sets of calculations were made, differing in the box size. 
   The numerical resolution was varied systematically, see 
   Table \ref{Calcu} for complete details. In
   the R--set the used box size was the standard $1 \times 2\pi \times
   4$ in units of the initial density scale height $H_{0}$, as used by
   e.g. Hawley et al. (\cite{hgbI}) and Brandenburg et
   al. (\cite{bnstI}) to be able to compare the results with previous
   studies and thereby validate the code. In the T--set the smallest
   box size in which the MRI was still seen to be active and where the 
   effects of the boundaries were not yet significant, namely $1 \times 1
   \times 2$, was chosen in order to achieve the best possible spatial
   resolution. The lowest resolution runs were advanced up to
   $\sim\!150$ orbits, whereas the largest calculation could only be
   continued until 13 orbits in a reasonable amount of time. The
   duration of the largest calculation determined the length of the
   time average for the energies and the $\alpha$--parameter. In the
   calculations we systematically increase the resolution keeping the
   box size and other parameters fixed, and calculate the resulting
   Shakura--Sunyaev $\alpha$--parameter in the following two ways. 
   Firstly, we calculate $\alpha$ using the Maxwell and Reynolds 
   stresses (see e.g. Brandenburg et al. \cite{bnstI})
   \begin{equation}
   \alpha_{\rm stress} = \frac{<\!\rho u_{x}u_{y}\!> - \mu_{0}^{-1} <\!B_{x}B_{y}\!>}{\frac{3}{2} \Omega_{0} c_{\rm s0} H_{0} <\!\rho_{0}\!>}\;.
   \label{Alpha1}
   \end{equation}    
   Secondly, we use the rate of dissipation
   \begin{equation}
   \alpha_{\rm diss} = \frac{(\frac{d}{dt}<\!\rho e\!> )_{\rm diss}}{(\frac{3}{2}\,\Omega_{0})^{2} c_{\rm s0} H_{0}<\!\rho_{0}\!>}\;.
   \label{Alpha2}
   \end{equation}

   In Eqs. (\ref{Alpha1}) and (\ref{Alpha2}) the square brackets
   denote volume averages, and we use the initial values of density,
   sound speed and density scale height as the normalisation
   factor. As the values of $\alpha$ from both methods coincide rather
   well, in the following we give only results from the first
   method. It is worth noting that the values of $\alpha$ calculated
   in this way are smaller than the values of e.g. Hawley et
   al. (\cite{hgbI}, \cite{hgbII}) by a factor of $3/\sqrt{2}$ due to
   different normalisation factors.

   \begin{figure*}
   \centering
   \includegraphics[width=0.9\textwidth]{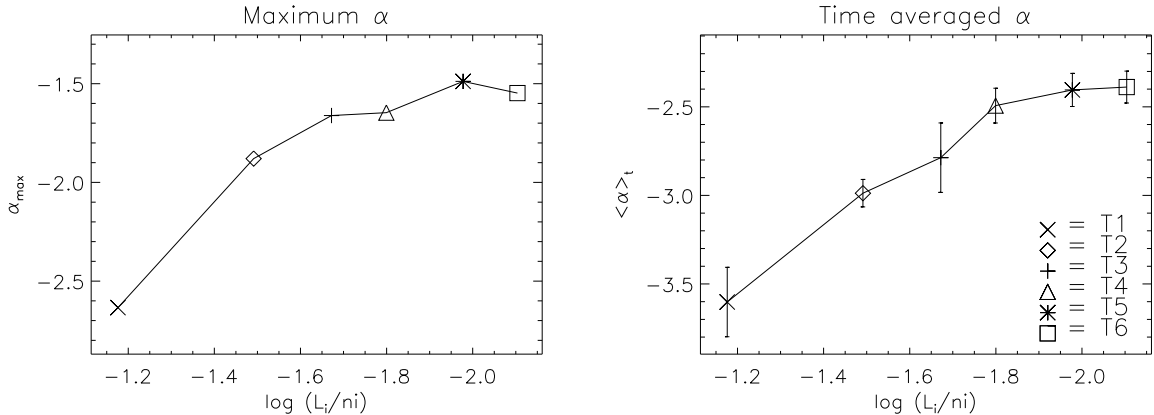}
   \vspace*{-6.cm}
      \caption{Time averages of the $\alpha$--parameter as a function of mesh size, $\Delta_{i}$, where $i = x, y,$ or $z$.}
       \label{Alpha}
       \vspace*{-0.3cm}
   \end{figure*}


\section{Results}
   The overall behaviour of the system is clearly visible from
   Fig.\,\ref{FigEne} where we plot the kinetic and magnetic energies
   for the run R1. The behaviour of the energies in the higher
   resolution simulations is very similar to the run R1, except that
   the energies are growing with resolution, see
   Fig.\,\ref{ResEne}. After the initial exponential growth, the
   energies saturate into a self--sustained turbulent state. Although
   large fluctuations are seen, no persistent decay or growth is seen
   after about five orbits. The initially weak magnetic field is
   amplified considerably due to dynamo action which leads into a
   configuration where the field is strongly dominated by the
   azimuthal component, followed by significantly weaker radial and
   vertical components. The behaviour of the magnetic energy is very
   similar to the earlier studies (e.g. Brandenburg et
   al. \cite{bnstI}). The kinetic energy, however, is slightly
   dominated by the radial component in our simulations, whereas in
   some earlier works the azimuthal component has also dominated the
   kinetic energy. Even though the magnetic energy is at all times
   larger than the kinetic energy, the system is dominated by thermal
   energy which is usually of the order of $\sim\!100$ times larger
   than the magnetic energy. This causes the ratio of thermal to
   magnetic pressure to be such that the MRI is active at all times in
   the simulations.

   Other characteristic numbers describing the system include the
   ratio of the Maxwell and Reynolds stresses, which are the main
   transport terms extracting energy from the Keplerian flow and
   transforming it into turbulent energy. In previous studies this
   ratio has been between 3 and 6. The value found in our simulations
   is close to 3, being in satisfactory agreement with the previous
   results. Although the results are not exactly identical to the
   earlier ones, the differences are not huge, and the basic
   characteristics can be well described by our model, indicating that
   it is valid.

   The maximum and time averaged energies as functions of the mesh
   size, $\Delta_{i} = L_{i}/ni$, for the T--set are shown in
   Fig.\,\ref{ResEne}. The maximum and time averaged kinetic energies
   both seem to continue almost linear growth beyond the resolutions
   investigated here, whereas the magnetic energies seem to saturate
   at the highest resolutions. A very similar trend was also found in
   the R--set (not shown), which indicates that the small box
   calculations are a valid tool in the study of resolution
   dependencies.

   In Fig.\,\ref{Alpha} the maximum and time averaged
   $\alpha$--parameters as functions of mesh size are shown for the
   T--set of runs. The actual numerical values of $\alpha$ are given
   in Table\,\ref{Calcu}. As can be seen from this table for the
   R--set, doubling the resolution (from R1 to R3), the value of
   $\alpha$ increases by a factor of $\sim\!3$, which is even more
   dramatic than in previous studies (e.g. Brandenburg et
   al. \cite{bnstII}).  In the T--set, the strong growth seen in the
   lowest resolutions seems to stop when resolution is increased
   further from the run T4. For the largest calculations, $\alpha$ is
   almost constant. The largest values of time averaged $\alpha$ we
   obtained were of the order of $0.004$ which are roughly two orders
   of magnitude too low to account for the aforementioned
   observational results of dwarf novae.

\vspace*{-0.4cm}
\section{Conclusions}
   In this letter, we have numerically investigated the resolution
   dependence of the Shakura \& Sunyaev $\alpha$--parameter from a
   local MHD model of a weakly magnetised accretion disk where the MRI
   is acting. We find that the numerical resolution affects the
   energies and the values of $\alpha$ significantly at lower
   resolutions. However, the strong growth of magnetic energy stops at
   the highest resolutions indicating that the Maxwell stress cannot
   extract more energy from the Keplerian shear flow when the
   resolution is increased which is also reflected in saturating trend
   seen in the values of $\alpha$. As the largest values of $\alpha$
   obtained in this study were only of the order $10^{-3}$, the
   obvious conclusion is that it is not possible to obtain high enough
   $\alpha$s due to the MRI, at least not from local restricted models
   as the one studied here. There are, however, recent global studies
   of the MRI (e.g. Hawley \cite{Hawley}), which models exhibit large
   scale wave phenomena producing significantly higher values of
   $\alpha$ (of the order $\sim\!0.1$). 

   Another hydromagnetic instability, namely the accretion ejection
   instability (AEI), investigated by Caunt \& Tagger (\cite{Suti01})
   has recently been identified as a source of angular momentum
   transport, producing similar physical characteristics (spiral waves
   and comparable values of $\alpha$) as the global MRI simulations,
   but which instability cannot operate without a fairly strong
   magnetic field. The AEI cannot generate this field by itself so it
   has to originate either from an external source or be generated by
   some intrinsic dynamo process, such as the one caused by the
   MRI. This suggests that the MRI is still a viable candidate to be
   responsible for the angular momentum transport (possibly indirectly
   via the AEI) in accretion disks and that it is the restricted
   geometry of the local models behind the too small values of
   $\alpha$.

\begin{acknowledgements}
      The calculations were carried out using the supercomputers
      hosted by \emph{CSC--Scientific Computing Ltd.}, Espoo,
      Finland. The authors wish to thank Prof. Axel Brandenburg for
      his valuable comments.
\end{acknowledgements}


\vspace*{-0.6cm}
\begin{thebibliography}{}
   \bibitem[1991]{balhaw1} Balbus, S. A., \& Hawley, J. F. 1991, ApJ,
     376, 214 

   \bibitem[1995]{bnstI}Brandenburg, A., Nordlund, \AA., Stein, R. F.,
   \& Torkelsson, U. 1995, ApJ, 446, 741

   \bibitem[1996]{bnstII}Brandenburg, A., Nordlund, \AA., Stein, R. F.,
   \& Torkelsson, U. 1996, ApJ Lett., 458, L45

   \bibitem[1993]{canniz} Cannizzo, J. K. 1993, ApJ, 333, 227   

   \bibitem[2001]{CauKo} Caunt, S. E. \& Korpi, M. J. 2001, A\&A, 369,
   706

   \bibitem[2001]{Suti01} Caunt. S. E. \& Tagger, M. 2001, A\&A, 367,
   1085

   \bibitem[1995]{hgbI} Hawley, J. F., Gammie, C. F., \& Balbus,
   S. A. 1995, ApJ, 440, 742

   \bibitem[1996]{hgbII} Hawley, J. F., Gammie, C. F., \& Balbus,
   S. A. 1995, ApJ, 464, 690

   \bibitem[2001]{Hawley} Hawley, J. F. 2001, ApJ, 554, 534

   \bibitem[2000]{millstone} Miller, K. A. \& Stone, J. M. 2000, ApJ,
   534,398

   \bibitem[1973]{shasun} Shakura, N. I., \& Sunyaev, R. A. 1973,
     A\&A, 24, 316

   \bibitem[1953]{spischwa} Spitzer, L, jr, \& Schwarzschild, M. 1953,
   ApJ, 118, 106

   \bibitem[1996]{stone} Stone, J. M., Hawley, J. F., Gammie, C. F.,
   \& Balbus, S. A. 1996, ApJ, 463, 656

\end{thebibliography}
\end{document}